\documentclass[nofootinbib,aps,reprint,superscriptaddress]{revtex4-1}
\usepackage{dcolumn}
\usepackage{bm}
\usepackage{amsmath}    
\usepackage{amsfonts} 
\usepackage{amssymb}
\usepackage{mathptmx}      
\usepackage{graphicx}   
\usepackage{subfigure}  
\usepackage{mathtools}
\usepackage{graphics}
\usepackage{epsfig}
\usepackage{tabularx}   
\usepackage{multirow}
\usepackage{bibentry}
\usepackage{braket}
\usepackage{algorithm}
\usepackage{algorithmic}
\usepackage{listings}
\usepackage{setspace}
\usepackage{fancyhdr}
\usepackage{slashed}
\usepackage{color}
\usepackage{xfrac}
\usepackage[pdftex, pdfborder= 0 0 0, citecolor=blue, urlcolor=blue, linkcolor=blue, colorlinks=true, bookmarksopen=true]{hyperref}

\def\Journal#1#2#3#4{{#1} {\bf#2}, #3 (#4)}

\def\NPA{{\rm Nucl. Phys.} A}
\def\NPB{{\rm Nucl. Phys.} B}
\def\PLB{{\rm Phys. Lett.}  B}

\def\PRD{{\rm Phys. Rev.} D}
\def\PRC{{\rm Phys. Rev.} C}

\def\IJMPA{{\rm Int. J. Mod. Phys.} A}

\def\la{\langle}
\def\ra{\rangle}

\def\be{\begin{equation}}
\def\ee{\end{equation}}
\def\bea{\begin{eqnarray}}
\def\eea{\end{eqnarray}}
\allowdisplaybreaks  

\newcommand{\pd}[2]{\frac{\partial{#1}}{\partial{#2}}}

\begin{document}\sloppy
\title{Variational Analysis of Mass Spectra and Decay Constants for Ground State Pseudoscalar and Vector Mesons in Light-Front Quark Model}
\author{Ho-Meoyng Choi}
\affiliation{Department of Physics, Teachers College, Kyungpook National University, Daegu, Korea 702-701}
\author{Chueng-Ryong Ji}
\affiliation{Department of Physics, North Carolina State University, Raleigh, North Carolina 27695-8202}
\author{Ziyue Li}
\affiliation{Department of Physics, North Carolina State University, Raleigh, North Carolina 27695-8202}
\author{Hui-Young Ryu}
\affiliation{Department of Physics, Teachers College, Kyungpook National University, Daegu, Korea 702-701}

\begin{abstract}
Using the variational principle, we compute mass spectra and decay constants of ground state pseudoscalar and vector mesons in the light-front quark model (LFQM) with the QCD-motivated effective Hamiltonian including the hyperfine interaction. By smearing out the Dirac delta function in the hyperfine interaction, we avoid the issue of negative infinity in applying the variational principle to the computation of meson mass spectra and provide analytic
expressions for the meson mass spectra. Our analysis with the smeared hyperfine interaction indicates that the interaction for the heavy meson sector including the bottom and charm quarks gets more point-like. We also consider the flavor mixing effect in our analysis and determine the mixing angles from the mass spectra of $(\omega,\phi)$ and $(\eta,\eta')$. Our variational analysis with the trial wave function including the two lowest order
harmonic oscillator basis functions appears to improve the agreement with the data of meson decay constants and the heavy meson mass spectra
over the previous computation handling the hyperfine interaction as perturbation.
\end{abstract}

\maketitle

\section{Introduction} 
\label{sec:Introduction}
Effective degrees of freedom to describe a strongly interacting system of hadrons
have been one of the key issues in understanding the non-perturbative nature of QCD in the
low energy regime. Within an impressive array of effective theories available nowadays,
the constituent quark model has been quite useful in providing a good physical picture of
hadrons just like the atomic model for the system of atoms.
Absorbing the complicated effect of quark, antiquark and gluon interactions into the effective constituent degrees of freedom, one may make the problem more tractable yet still keep some key features of the underlying QCD to provide useful predictions \cite{ConstituentQuarkfromQCD}.
The effective potentials used in constituent quark models are typically described by the flux tube configurations generated by the gluon fields as well as the effective ``one-gluon-exchange" calculation in QCD \cite{BoundStatesofQuarks1991,OneGluonExchange}.
In the QCD-motivated effective Hamiltonian, a proper way of dealing with the relativistic effects in the hadron system is quite essential
due to the nature of strong interactions. In particular, proper care and handling of relativistic effects has been emphasized in describing the hadrons made of $u$, $d$, and $s$ quarks and antiquarks.

As a proper way of handling relativistic effects, the light-front quark model (LFQM)~\cite{Kon,Cheng97,Hwang10,CCP,Card95}
appears to be one of the most efficient and effective tools in hadron physics as it takes advantage of the distinguished features of the light-front dynamics (LFD)~\cite{DiracForms,BPP}.
In particular, the LFD carries the maximum number~(seven) of the kinetic
(or interaction independent) generators and thus the less effort in dynamics is
necessary in order to get the QCD solutions that reflect the full Poincar$\acute{e}$ symmetries.
Moreover, the rational energy-momentum dispersion relation  of LFD, namely
$p^-=({\bf p}^2_\perp + m^2) / p^+$, yields the sign correlation between the light-front (LF) energy
$p^-(=p^0-p^3)$ and the LF longitudinal momentum $p^+(=p^0 + p^3)$ and
leads to the suppression of quantum fluctuations of the vacuum, sweeping
the complicated vacuum fluctuations into the zero-modes in the limit of
$p^+ \rightarrow 0$~\cite{Zero1,Zero2,Zero3}.
This simplification is a remarkable advantage in LFD and facilitates the partonic interpretation of the amplitudes.
Based on the advantages of the LFD, the LFQM  has been developed~\cite{CJ_99} and subsequently applied for
various meson phenomenologies such as the mass spectra of both heavy and light mesons~\cite{CJ_Bc},
the decay constants, distribution amplitudes, form factors and generalized parton distributions~\cite{BPP,CJ_99,CJ_Bc,CJ_DA,Jaus99,Jaus03,Cheng04,CJ_PV,MF12,Jaus90,Choi07}.

Despite these successes in reproducing the general features
of the data, however, it has proved very difficult to obtain direct connection
between the LFQM and QCD.
Typically, rigorous derivations of the connection between the effective constituent degrees of freedom and
the fundamental QCD quark, antiquark and gluon degrees of freedom have been explored by solving momentum-dependent
mass gap equations as discussed in many-body Hamiltonian approach~\cite{many-body-H}, Dyson-Schwinger approach~\cite{DS}, etc.
Although one has not yet explored solving the momentum-dependent mass gap equation in LFD,
there has been some attempt to derive an effective LF Hamiltonian starting from QCD using the discrete light-cone quantization
(DLCQ) and solve the corresponding equation of motion approximately for the quark and antiquark bound-states to provide
semianalytical expressions for the masses of pseudoscalar and vector mesons~\cite{Chris-Pauli}.
The attempt to link between QCD and LFQM is also supported by our recent analyses of quark-antiquark distribution amplitudes
for pseudoscalar and vector mesons in LFQM~\cite{CJ_V14},
where we presented a self-consistent covariant description of  twist 2 and twist 3
quark-antiquark distribution amplitudes for pseudoscalar and vector mesons in LFQM
to discuss the link between the chiral symmetry of QCD and the LFQM.
Our results for the pseudoscalar and vector mesons~\cite{CJ_V14} effectively indicated that the constituent quark and
antiquark in the LFQM could be considered as the dressed constituents including the zero-mode quantum fluctuations
from the QCD vacuum.
Moreover, the light-front holography based on the 5-dimensional anti-de Sitter (AdS) spacetime and the conformal
symmetry has given insight into the nature of the effective confinement potential and the resulting light front wavefunctions for both
light and heavy mesons~\cite{Brodsky2006}.
As we have shown in Ref.~\cite{ChoiJi2008}, our LFQM analysis of the pion form factor provided compatible results
both in spacelike and timelike regions with the holographic approach to LF QCD~\cite{Brodsky2008}.
These developments motivate our present work for the more-in-depth analysis of
the mass spectra and decay constants for the ground state pseudoscalar and vector mesons in LFQM.

In LFQM, the LF wave function is independent of all reference frames related by the front-form boosts
because the longitudinal boost operator as well as the LF transverse boost operators are all kinematical.
This is clearly an advantageous feature unique to LFQM, which makes the calculation of observables such as
mass spectra, decay constants, form factors, etc. much more effective.
Computing the meson mass spectra, however, we have previously~\cite{CJ_99,CJ_Bc} treated the hyperfine interaction
as a perturbation rather than including it in the variation procedure to avoid the negative infinity from the Dirac delta function
contained in the hyperfine interaction.
In the present work, we smear out the Dirac delta function by a Gaussian distribution and
resolve the infinity problem when variational principle is applied to the hyperfine interaction. We obtain optimal model parameters
in our variational analysis including the hyperfine interaction and examine if it improves phenomenologically our numerical results compared
to the ones obtained by the perturbative treatment of the hyperfine interaction. For our trial wave function, we also
take a larger harmonic oscillator (HO) basis to see if it provides any phenomenological improvement in our predictions of
mass spectra and decay constants for ground state pseudoscalar and vector mesons.

The paper is organized as follows: In Sec.~\ref{sec:ModelDescription}, we describe our QCD-motivated effective Hamiltonian with the smeared-out hyperfine interaction.
Using
the mixture of the two lowest order HO states
as our trial wave function of the variational principle,
we find the analytic formula of the mass eigenvalues for the ground state pseudoscalar and vector mesons.
The optimum values of model parameters are also presented in this section.
In Sec.~\ref{sec:ResultsandDiscussion},
we present our numerical results of the mass spectra
obtained by taking a larger HO basis in the trial wave function
and compare them with the experimental data as well as our previous calculations~\cite{CJ_99,CJ_Bc}.
To test our trial wave function with the parameters obtained from the variational principle,  we also calculate the meson decay constants and compare them with the experimental data as well as other available theoretical predictions.
Summary and conclusion follow in Sec.~\ref{sec:SummaryandConclusion}.
The detailed procedure of fixing our parameters through variational principle is presented in Appendix \ref{sec:AppendixFixingParameters}.

\section{Model Description} 
\label{sec:ModelDescription}
As mentioned in the introduction, there has been an attempt to derive an effective LF Hamiltonian starting from QCD
using DLCQ~\cite{Chris-Pauli}. Transforming the LFD variables to the ordinary variables in the instant form dynamics (IFD),
one may see the equivalence between the resulting effective LF Hamiltonian for the quark and antiquark bound-states
and the usual relativistic constituent quark model Hamiltonian for mesons typically given in the rest frame of the meson, i.e.
the center of mass (C.M.) frame for the constituent quark and antiquark system. It may be more intuitive to express the
effective LF Hamiltonian describing the relativistic constituent quark model system for mesons in terms of the ordinary IFD variables.
Effectively, the meson system at rest is then described as an interacting bound system of effectively dressed valence quark and antiquark
typically given by the following QCD-motivated effective Hamiltonian in the quark and antiquark C.M. frame~\cite{CJ_99,CJ_Bc}:
  \begin{align}\label{eqn:H}
    &H_{\rm C.M.}= \sqrt{m_{q}^{2}+\vec{ k
    }^{2}}+\sqrt{m_{\bar{q}}^{2}+\vec{k}^{2}}+V,
  \end{align}
where $\vec{k}=( \mathbf{k_{\bot}},k_{z} )$ is the relativistic three-momentum of the constituent quarks and $V$ is the effective potential
between quark and antiquark in the rest frame of the meson. The effective potential $V$ is typically given by
the linear confining potential $V_{\text{conf}}$ plus the effective one-gluon-exchange potential $V_{\text{oge}}$.
For $S$-wave pseudoscalar and vector mesons,
the effective one-gluon-exchange potential reduces to the coulomb potential $V_{\text{coul}}$
plus the hyperfine interaction $V_{\text{hyp}}$.
Thus, one may summarize $V$ as
\begin{align}\label{poten}
  V&=V_{\text{conf}}+V_{\text{oge}} \nonumber \\
  &=\underbrace{ a \ +\ b\ r
  }_{\text{conf}}\ \underbrace{\overbrace{-\
    \frac{4\alpha_{s}}{3r}}^{\text{coul}}\
  +\
  \overbrace{ \frac{2}{3}\frac{\mathbf{S}_{q}\cdot\mathbf{S}_{\bar{q}}}{m_{q}m_{\bar{q}}}\nabla^{2}V_{\text{coul}} }^{\text{hyp}} }_{\text{oge}},
\end{align}
where $\alpha_s$ is the strong interaction coupling constant~\footnote{Although one may consider a running coupling constant, we take $\alpha_s$ as one of the variation parameters in this work.}, $\langle\mathbf{S}_{q}\cdot\mathbf{S}_{\bar{q}} \rangle=1/4\ (-3/4)$ for the vector (pseudoscalar) meson and $\nabla^{2}V_{\text{coul}}=(16\pi\alpha_{s}/3) \delta ^{ 3 }(\mathbf{r})$.
Reduction of the LF Hamiltonian in QCD to a similar form of the effective Hamiltonian in the C.M.
frame of the quark and antiquark system given by Eqs. (\ref{eqn:H}) and (\ref{poten}) was discussed in Ref.~\cite{Chris-Pauli}.
For the hyperfine interaction $V_{\text{hyp}}$, one may consider the relativization such as $V_{\text{hyp}}
\rightarrow \sqrt{m_q m_{\bar q}/E_q E_{\bar q}} V_{\text{hyp}} \sqrt{m_q m_{\bar q}/E_q E_{\bar q}}$~\cite{IsgurCapstickBaryon,GodfreyIsgurMeson}.
Such relativization may be important for the $\delta ^{ 3 }(\mathbf{r})$-type potential without any smearing in computing
particularly the light meson sector. Since we apply the variational principle even for the hyperfine interaction in this work smearing
out the Dirac delta function to resolve the infinity problem, we naturally introduce a smearing parameter which may effectively compensate
the factor due to the relativization. With this treatment, we are able to provide explicit analytic expressions for the meson mass spectra
(see Eq.(\ref{eqn:MassEquationMix})).

While the effective bound-state mass square $M^2_{q \bar{q}} $ is given by $M^2_{q \bar{q}} = (P^0_{\rm C.M.})^2$ in the C.M. frame of the constituent quark and antiquark system, the energy-momentum dispersion relation in LFD is given by $M^2_{q \bar{q}} = P^+P^- - \mathbf{P}_{\bot}^{2}$, where the four-momentum of the bound system is denoted by $P^{\mu}=(P^{+},P^{-},\mathbf{P}_{\bot})=(P^{0}+P^{3},P^{0}-P^{3},\mathbf{P}_{\bot})$. From this, one may consider the LFQM mass square operator ${\hat P}^+{\hat P}^- - \mathbf{{\hat P}}_{\bot}^{2}$
(that provides the eigenvalues $ P^+P^- - \mathbf{P}_{\bot}^2$) as the square of the effective Hamiltonian given by Eq. (\ref{eqn:H}), i.e. $H_{\rm C.M.}^2$. Since the eigenvalues and the expectation values are same for the eigenstates, we compute the expectation value $\langle H_{\rm C.M.}\rangle$
using the variation principle. Alternatively, one may consider computing the expectation value  $\langle H_{\rm C.M.}^2\rangle$ in view of
the LFQM mass square operator ${\hat P}^+{\hat P}^- - \mathbf{{\hat P}}_{\bot}^{2}$ being $\langle H_{\rm C.M.}^2\rangle$.  Although
$\langle (\Delta H_{\rm C.M.})^2\rangle=0$ in principle for the eigenstates, it may be interesting to examine numerically how small the corresponding
deviation $\langle (\Delta H_{\rm C.M.})^2\rangle = \langle H_{\rm C.M.}^2\rangle - \langle H_{\rm C.M.} \rangle^2$ is.
More future works complementary to our present computation
of $\langle H_{\rm C.M.} \rangle$ can be suggested in variational analysis. In this work, we examine the $\chi^2$ values of our computational
results in comparison with experimental data to get optimal parameter values in the $\langle H_{\rm C.M.} \rangle$ computation.
This will provide useful ground information for any alternative and/or further works beyond the present analysis.

As discussed earlier, the longitudinal boost operator as well as the LF transverse boost operators are all kinematical and
thus the LF wave function does not depend on the external momentum, i.e. $P^+$ and $\mathbf{P}_{\bot}$.
In effect, the determination of the LF wave function in the meson rest frame such as $P^+=M_{q \bar{q}}$ and $\mathbf{P}_{\bot}=0$ won't hinder its use for any other values of $P^+$ and $\mathbf{P}_{\bot}$.
This provides the applicability of LFQM for the computation of observables beyond the meson mass spectra.

The wave function is thus represented by the Lorentz invariant internal variables $x_{i}=p_{i}^{+}/P^{+},\  \mathbf{k}_{\bot i}=\mathbf{p}_{\bot i}-x_{i}\mathbf{P}_{\bot}$ and helicity $\lambda_{i}$, where $p^\mu_i$ is the momenta of constituent quarks.
Explicitly, the LF wave function of the ground state mesons is given by
\begin{align}
  \Psi_{100}^{JJ_z}(x_{i},\mathbf{k}_{\bot
  i},\lambda_{i})=\mathcal{R}_{\lambda_{q}\lambda_{\bar{q}}}^{JJ_z}(x_{i},\mathbf{k}_{\bot
  i})\Phi(x_{i},\mathbf{k}_{\bot i}),
  \label{eqn:Wavefunction}
\end{align}
where $\Phi$ is the radial wave function and $\mathcal{R}_{\lambda_{q}\lambda_{\bar{q}}}^{JJ_z}$ is the interaction-independent spin-orbit wave function.
The spin-orbit wave functions for pseudoscalar and vector mesons are given by~\cite{CJ_99,Jaus91}
\begin{align}
  \begin{split}
    \mathcal{R}_{\lambda_{q}\lambda_{\bar{q}}}^{00}&
    =\frac{-\bar{u}_{\lambda_{q}}(p_q)\gamma_{5}\nu_{\lambda_{\bar{q}}}(p_{\bar{q}})}{\sqrt{2}\sqrt{M_{0}^{2}-(m_{q}-m_{\bar{q}})^{2}}},
    \\
    \mathcal{R}_{\lambda_{q}\lambda_{\bar{q}}}^{1
      J_{z}}&=\frac{-\bar{u}_{\lambda_{q}}(p_{q})\left[\slashed
	\epsilon(J_{z})-\frac{\epsilon \cdot
	  (p_{q}-p_{\bar{q}})}{M_{0}+m_{q}+m_{\bar{q}}}\right]
	  \nu_{\lambda_{\bar{q}}}(p_{\bar{q}})}{\sqrt{2} \sqrt{M_{0}^{2}-(m_{q}-m_{\bar{q}})^{2}}},
  \end{split}
\end{align}
where $\epsilon^{\mu}(J_{z})$ is the polarization vector of the vector meson and
the boost invariant meson mass squared $M_{0}^{2}$ obtained from the free energies of the constituents is given by
\begin{align}
  M_{0}^{2}=\frac{\mathbf{k}_{\bot}^{2}+m_{q}^{2}}{x}+\frac{\mathbf{k}_{\bot}^{2}+m_{\bar{q}}^{2}}{1-x}.
\end{align}
The spin-orbit wave functions satisfy the relation $\sum_{\lambda_{q}\lambda_{\bar{q}}} \mathcal{R}_{\lambda_{q}\lambda_{\bar{q}}}^{JJ_{z}\dagger} \mathcal{R}_{\lambda_{q}\lambda_{\bar{q}}}^{J J_{z}}=1$ for both pseudoscalar and vector mesons.

To use a variational principle, we take our trial wave function as an expansion of the true wave function in the HO basis.
We use the same trial wave function
expanded with the two lowest order HO wave functions
$\Phi=\sum_{n=1}^{2} c_n \mathcal{\phi}_{nS}$
for both pseudoscalar and vector mesons, where
  \begin{align}\label{subeqn:radial}
    &\mathcal{\phi}_{1S}(x_{i},\mathbf{k}_{\bot
  i})=\frac{4\pi^{3/4}}{\beta^{3/2}}\sqrt{\frac{\partial
    k_{z}}{\partial x}}\;e^{-\frac{\vec{k}^{2}}{2\beta^{2}}},
  \end{align}
\begin{align}\label{subeqn:2S}
    \begin{split}
    \mathcal{\phi}&_{2S}(x_{i},\mathbf{k}_{\bot i})
    =\frac{4\pi^{3/4}}{\sqrt{6}\beta ^{7/2}}\left(2
    \vec{k}^2-3 \beta ^2\right)\sqrt{\frac{\partial
    k_{z}}{\partial x}}\;e^{-\frac{\vec{k}^2}{2\beta ^2}},
    \end{split}
\end{align}
and $\beta$ is the variational parameter.
 We should note here that our LF wave functions $\phi_{nS}$ are dependent on $M^2_0$ and thus cannot be factorized into
 a function of $\mathbf{k}_{\bot i}$ multiplied by another function of $x_{i}$. In particular, $\vec{k}^2$ in Eqs.~(\ref{subeqn:radial})
 and (\ref{subeqn:2S}) is given by $\vec{k}^2={\bf k}^2_\perp + k^2_z$ where
$k_{z}=(x-1/2)M_{0}+(m_{\bar{q}}^{2}-m_{q}^{2})/2M_{0}$.
For instance, $e^{-\vec{k}^2/2\beta ^2}=e^{m^2/2\beta^2} e^{-M^2_0/8\beta^2}$ in the case of equal quark and antiquark mass $m_q=m_{\bar q}=m$.
The variable transformation
$(x, \mathbf{k}_{\bot})\rightarrow \vec{k}=(\mathbf{k}_{\bot},k_{z})$ requires the Jacobian factor
given by $ \partial k_{z}/\partial x=M_{0} [1- (m_q^2 - m_{\bar q}^2)^2 / M_0^4 ]/4x(1-x)$ as one can see from Eqs.~(\ref{subeqn:radial})
and (\ref{subeqn:2S}). The normalization of the wave function $\mathbf{\phi}_{nS}$ is thus given by
\begin{align}
\int_{0}^{1}dx\int
  \frac{d^{2}\mathbf{k}_{\bot}}{16\pi^{3}}
  |\phi_{nS}(x_{i},\mathbf{k}_{\bot i})|^{2}=1.
\end{align}

With $\Phi=\sum_{n=1}^{2} c_n \mathcal{\phi}_{nS}$,
we evaluate the expectation value of
the Hamiltonian in Eq.~(\ref{eqn:H}), i.e.  $\langle \Phi|H_{\rm C.M.}|\Phi\rangle$ which depends on the variational parameter $\beta$.
According to the variational principle, we can set the upper limit of the ground state's energy by calculating the expectation value of
the system's Hamiltonian with a trial wave function.
In our previous calculations~\cite{CJ_99,CJ_Bc}, which we call ``CJ model", we first evaluate the expectation value of the central Hamiltonian $T + V_{\rm conf}+V_{\rm coul}$
with the trial function $\phi_{1S}$,
where $T$ is the kinetic energy part of the Hamiltonian.
Once the model parameters are fixed by minimizing the expectation value
$\langle\phi_{1S}|(T + V_{\rm conf}+V_{\rm coul})|\phi_{1S}\rangle$,
then the mass eigenvalue of each meson is obtained as
$M_{q\bar{q}}= \langle\phi_{1S}|H_{\rm C.M.}|\phi_{1S}\rangle$.
The hyperfine interaction $V_{\rm hyp}$ in CJ model, which contains a Dirac delta function, was treated as perturbation to the Hamiltonian and was left out in the variational process that optimizes the model parameters.
The main reason for doing this was to avoid the negative infinity generated by the delta function as was pointed out in~\cite{IsgurCapstickBaryon}.
Specifically, $\langle\phi_{1S}|V_{\rm hyp}|\phi_{1S}\rangle$
for pseudoscalar mesons
decreases faster than other terms that increase as $\beta$ increases
and the expectation value of the Hamiltonian is unbounded from below.

The singular nature of the hyperfine interaction and its regularization is a standard topic in atomic physics
and the atomic analysis has been carried out to extraordinary precision~\cite{Brodsky1977}.
In particular, a Bethe-Salpeter based bound-state formalism was applied to the calculation of recoil contributions of order
$m\alpha^6$ to hyperfine splitting in ground-state positronium~\cite{Adkins1998}.
Instead of dropping the relative energy dependence in favor of equations with a simpler kinematical structure but a more complicated effective kernel,
the Barbieri-Remiddi formalism~\cite{Barbieri1978} was discussed as an effective way to handle significant complications concerning
the Bethe logarithm~\cite{Bethe1947}. As discussed in Ref.~\cite{Adkins1998}, the $\delta$ function of the relative energy $p_0$ is replaced
by a smearing function of $p_0$ in the Barbieri-Remiddi formalism~\cite{Barbieri1978}.
In LFD, the equal LF time $x^+(=x^0+x^3)$ correlates the ordinary time $x^0$ and space $x^3$ so that
the idea of smearing $p^0$ in the Barbieri-Remiddi formalism may be extended to smear
the $\delta^{ 3 }(\mathbf{r})$ function in hyperfine interaction discussed in the present work.
In this respect, our regularization procedure discussed below would also be valid and is compatible for the
hyperfine splitting in atoms. Analytic treatment of positronium spin splittings was presented in LF QED~\cite{Jones1997}
and more recent DLCQ application to the analysis of $\mu^+ \mu^-$ bound state spectrum can be found in Ref.~\cite{Lamm2014}.

To avoid the negative infinity, we thus use a Gaussian smearing function to weaken the singularity of $\delta ^{3}(\mathbf{r})$ in hyperfine interaction, viz.~\cite{IsgurCapstickBaryon,GodfreyIsgurMeson},
  $\delta^{3} (\mathbf{r})\rightarrow
  (\sigma^3/\pi^{3/2})e^{-\sigma^{2}\mathbf{r}^{2}}$.
Once the delta function is smeared out like this,
a true minimum for the mass occurs at a finite value of
$\beta$.
The analytic formulae of mass eigenvalues for our modified Hamiltonian with the smeared-out hyperfine interaction, i.e. $M_{q\bar{q}}=\langle \Phi |H_{\rm C.M.}|\Phi \rangle$, are found as follows:~\footnote{Although the true minimum occurs with the smeared-out hyperfine
interaction even for $\phi_{1S}$ case, we found that the phenomenological results do not show any significant improvement compared to CJ model.}
\begin{align} \label{eqn:MassEquationMix}
    \begin{split}
M_{q \bar{q}}&= a + \frac{b}{\beta\sqrt{\pi}}
\left( 3 - c_{1}^{2} -  2\sqrt{\frac{2}{3}}c_1 c_2 \right)\\
 &   +\frac{\beta}{\sqrt{\pi }}  \sum_{i=q,{\bar q}}
\left\{
\sqrt{\pi } \left(\sqrt{6}c_1 c_2 -3c_{2}^{2}\right)
U\left(-\frac{1}{2},-2, z_i\right)
\right.
 \\
 &+\frac{1}{3} c_{2}^{2} z_i^2   e^{\frac{z_i}{2}} \left(3 - z_i \right) K_2\left(\frac{z_i}{2}\right)
 \\
 &\left.+ \frac{1}{6}z_i  e^{\frac{z_i}{2}}
\left( 2 c_{2}^{2} z_i^2 - 3  c_{1}^{2} - 6 \sqrt{6}c_1 c_2  + 9 \right) K_1\left(\frac{z_i}{2}\right)
\right\}
 \\
 &   - \frac{4 \alpha _s \beta}{9\sqrt{\pi}}
\left\{
 5  +  c_{1}^{2} + 6 \sqrt{2/3}c_1c_2 \right.\\
	&-	\frac{4 \beta ^2 \sigma ^3 \langle
   \mathbf{S}_{q}\cdot \mathbf{S}_{\bar{q}}\rangle}{ \left(\beta ^2+\sigma ^2\right)^{7/2} m_q m_{\bar{q}}} \left[\left(2 \sqrt{6}c_1c_2 +3-c_{1}^{2}\right) \sigma^4\right.\\
	&\left.\left.+2 \beta ^2 \left(2 c_{1}^{2} + \sqrt{6} c_1 c_2\right) \sigma^2+2 \beta^4\right]\right\},
    \end{split}
\end{align}
where $z_i = m^2_i /\beta^2$ and $K_{1}$ is the modified Bessel function of the second kind and $U(a,b,z)$ is Tricomi's (confluent hypergeometric) function.
We should note that the mass formula for the delta-function hyperfine interaction
corresponds to Eq.~(\ref{eqn:MassEquationMix}) in the limit of $\sigma\to\infty$.
We then apply the variational  principle, i.e. $\partial M_{q\bar{q}}/\partial\beta=0$, to find the optimal model parameters in order to get a best fit for the mass spectra of ground state pseudoscalar and vector mesons (a more detailed description of this procedure can be found in~ \ref{sec:AppendixFixingParameters}).

Our optimized potential parameters are obtained as
$\{ a=-0.6699 \ \text{GeV}, b=0.18 \ \text{GeV}^{2}, \alpha_{s}=0.4829 \}$.
For the  best fit of the ground state mass spectra,
we obtain
$c_{1}=+\sqrt{0.7}$ and $c_2=+\sqrt{0.3}$.
We should note that our potential parameters are quite comparable with the ones suggested by Scora and Isgur~\cite{ScoraIsgur1995}, where they obtained $a=-0.81$ GeV, $b=0.18\ \text{GeV}^{2}$, and $\alpha_{s}=0.3 \sim 0.6$.
For a comparison, the coupling constant we found in our previous
CJ model~\cite{CJ_99,CJ_Bc} was $\alpha_{s}=0.31$.

While we use the common potential parameters $(a, b,\alpha_s)$ for all the mesons,
it was shown in~\cite{GodfreyIsgurMeson,BBD} that if a smearing procedure for the $\delta ^{3}(\mathbf{r})$ function
is used, then a large Gaussian parameter $\sigma$ is obtained for the heavy quark sector.
In our updated potential model using the smeared hyperfine interaction $(\sigma^3/\pi^{3/2})e^{-\sigma^{2}\mathbf{r}^{2}}$,
we also confirm the same observation as in~\cite{GodfreyIsgurMeson} for the heavy meson sector including $(b, c)$ quarks.
Thus, we differentiate the smearing parameter $\sigma$ for the heavy ($b,c$) sectors such as $(c\bar{c}, b\bar{c}, b\bar{b})$
from the other ($q\bar{q}$) sectors by introducing multiplicative factor in front of
$\sigma$, i.e. $\sigma\to\lambda\sigma$ with $\lambda>1$, while other potential
parameters $(a, b, \alpha_s)$ remain the same for all $(q\bar{q})$
sectors. This differentiation is to accommodate the hyperfine splittings for the heavy $(b,c)$ quark sectors
as we will show in the next section. Our new updated results with $\lambda$ differentiation using the common potential parameters $(a, b, \alpha_s,\sigma)$
for all meson sectors show definite improvement in the $\chi^2$ fit of the experimental data for meson masses.

Our optimal constituent quark masses and the smearing parameters $\sigma$  are listed in Table~\ref{tab:quarkmass}.
Since we included the hyperfine interaction with smearing function entirely in our variational process, we now obtain the two different
sets of $\beta$ values, one for pseudoscalar and the other for vector mesons, respectively.
The optimal Gaussian parameters $\beta_{q\bar{q}}$ for pseudoscalar and vector mesons are also
listed in Table~\ref{tab:betaP}.
We should note that the values of the
multiplicative factor $\lambda$ to get the best fits for
the mass eigenvalues
are obtained as $\lambda=(2, 2.3, 3)$ for
$(c\bar{c}, b\bar{c}, b\bar{b})$ sectors.
As a sensitivity  check, however, we present the numerical results  with the following theoretical error bars
 $\lambda=(2^{+1}_{-1}, 2.3^{+1}_{-1}, 3^{+2}_{-2})$ for
$(c\bar{c}, b\bar{c}, b\bar{b})$ sectors, respectively.
Although one may fine-tune more to improve the hyperfine
splittings for the heavy-light sectors by using different set of $\lambda$ parameters,
we set $\lambda=1$ for any other $q\bar{q}$ sectors except
$(c\bar{c}, b\bar{c}, b\bar{b})$ sectors in this work for simplicity.
\begin{table}[h]
  \caption{\label{tab:quarkmass}Constituent quark masses [GeV] and
  the smearing parameter $\sigma$ [GeV]
  obtained by the variational principle for the Hamiltonian with a
  smeared-out hyperfine interaction. Here $\text{q}=\text{u}\; \text{and}\;
d$.}
    \begin{tabular*}{\columnwidth}{l@{\extracolsep{\fill}}cccc}
				\hline
      $m_{\text{q}}$ & $m_{\text{s}}$ & $m_{\text{c}}$ & $m_{\text{b}}$
      & $\sigma$
      \\
      \hline
      0.205 & 0.380 & 1.75 & 5.15 & 0.423\\
			\hline
    \end{tabular*}
\end{table}

\begin{table*}[t]
  \caption{\label{tab:betaP}The Gaussian parameter $\beta$ [GeV]
  for ground state pseudoscalar ($J^{PC}=0^{-+}$) and vector ($1^{--}$)
  mesons obtained by the variational principle. $\text{q}=\text{u}\;
  \text{and}\; \text{d}$.
We should note that $\lambda=(2^{+1}_{-1}, 2.3^{+1}_{-1}, 3^{+2}_{-2})$
are used to get $(\beta_{\text{cc}}, \beta_{\text{bc}}, \beta_{\text{bb}})$ values
and $\lambda=1$ is used to get
the rest of $\beta_{\text{qq}}$ values.}
    \begin{tabular*}{\textwidth}{l@{\extracolsep{\fill}}cccccccccc}
				\hline
    $J^{PC}$ &
    $\beta_{\text{qq}}$ & $\beta_{\text{qs}}$ & $\beta _{\text{ss}}$ & $\beta _{\text{qc}}$ &
  $\beta _{\text{cs}}$   & $\beta_{\text{cc}}$ & $\beta_{\text{qb}}$ & $\beta _{\text{bs}}$ & $\beta _{\text{bc}}$ &
  $\beta _{\text{bb}}$ \\
    \hline
     $0^{-+}$ &
    0.4465 & 0.3759 & 0.3445 & 0.3801 & 0.3859 & $0.5270^{+0.0291}_{-0.0235}$ &
    0.4226 & 0.4412 & $0.6646^{+0.0219}_{-0.0174}$ & $0.9906^{+0.0420}_{-0.0223}$\\
    \hline
         $1^{--}$ &
    0.2346 & 0.2598 & 0.2820 & 0.3445 & 0.3667 & $0.4914^{-0.0062}_{+0.0072}$ &
    0.4057 & 0.4321 & $0.6365^{-0.0058}_{+0.0056}$ & $0.9603^{-0.0122}_{+0.0075}$\\
		\hline
    \end{tabular*}
\end{table*}

We also determine the mixing angles from the mass spectra of $(\omega,\phi)$ and $(\eta,\eta')$.
Identifying $({\cal F,F'})=(\phi,\omega)$ and $(\eta,\eta')$ for vector and pseudoscalar nonets,
the flavor assignment of ${\cal F}$ and ${\cal F'}$ mesons in the quark-flavor basis
$n\bar{n}=(u\bar{u}+d\bar{d})/\sqrt{2}$ and
 $s\bar{s}$ is given by~\cite{FKS,OZI,Leut98}
  \begin{eqnarray}\label{Eq:mixang}
 \left( \begin{array}{cc}
 {\cal F}\\
 {\cal F'}
 \end{array}\,\right)
 =\left( \begin{array}{cc}
 \cos\alpha\;\; -\sin\alpha\\
 \sin\alpha\;\;\;\;\;\cos\alpha
 \end{array}\,\right)\left( \begin{array}{c}
 n\bar{n}\\
 s\bar{s}
 \end{array}\,\right)
 = U(\alpha)\left( \begin{array}{c}
 n\bar{n}\\
 s\bar{s}
 \end{array}\,\right),
 \end{eqnarray}
where $\alpha$ is the mixing angle in the quark-flavor basis.
For the $\eta-\eta'$ mixing, the SU(3) mixing angle $\theta$ in the
the flavor SU(3) octet-singlet basis $(\eta_8,\eta_1)$ can also be
used and the relation between the mixing angles is given by
$\theta=\alpha-{\rm arctan}\sqrt{2}\simeq\alpha - 54.7^{\circ}$\cite{PDG2014}.
Taking into account SU(3) symmetry breaking and using the parametrization for the
$(\rm mass)^2$ matrix suggested by Scadron~\cite{Scad}, we obtain~\cite{CJ_99}
\be\label{Eq:mixM}
\tan^2\alpha = \frac{(M^2_{{\cal F'}} - M^2_{s\bar{s}})(M^2_{\cal F} - M^2_{n\bar{n}})}
{(M^2_{\cal F'} - M^2_{n\bar{n}})(M^2_{s\bar{s}}-M^2_{\cal F})},
\ee
which is the model-independent equation for any $q\bar{q}$ meson nonets.
The details of obtaining
meson mixing angles using quark-annihilation diagrams are summarized in~\cite{CJ_99},
where the mixing angle $\delta=\alpha - 90^{\circ}$ is used in the quark-flavor basis.
In order to predict the $\omega-\phi$ and $\eta-\eta'$ mixing angles, we use the experimental
values of
$M_{\cal F} = (M_\phi, M_\eta)$ and $M_{\cal F'}=(M_\omega, M_{\eta'})$
as well as the masses of $M^V_{n\bar{n}}\;[M^V_{s\bar{s}}]=780\;(901)$ MeV
and $M^P_{n\bar{n}}\;[M^P_{s\bar{s}}]=140\;(726)$ MeV obtained from
$\la\Phi|H_{s\bar{s}}|\Phi\ra$ for both vector ($V$) and pseudoscalar ($P$) mesons, respectively.
Our prediction for $\omega-\phi$ mixing angle is
$\alpha_{\omega-\phi}=84.8^\circ$, which is about $5.2^\circ$ deviated from
the ideal mixing $\alpha^{\omega-\phi}_{\rm ideal}=90^\circ$. Our prediction for
$\eta-\eta'$ mixing angle is $\alpha_{\eta-\eta'}=36.3^\circ$,  which is in
agreement with the
range $34.7^\circ$ to $44.7^\circ$ of phenomenological values~\cite{FKS,PDG2014}.

Our updated model with the smeared hyperfine interaction appears to improve the result of mass spectrum, which is presented in the next section.
This may suggest that when using constituent quark models, the contact interactions has to be smeared out in general.
In fact, we think this smeared interaction is more consistent with the physical picture for a system of the effective constituent quarks which are not point-like.

For practical application of our model, we also compute the decay constants for the ground state pseudoscalar and vector mesons.
The decay constants are typically defined by
\begin{align}
  \begin{split}
    \langle0|\bar{q}\gamma^{\mu}\gamma_{5}q|P\rangle&=if_{P}P^{\mu},
    \\
    \langle0|\bar{q}\gamma^{\mu}q|V(P,h)\rangle&=f_{V}M_{V}\epsilon^{\mu}(h),
  \end{split}
\end{align}
for pseudoscalar and vector mesons, respectively.~
For the $\eta$ and $\eta'$ case, one may also define decay constants through
matrix elements of octet and singlet axial-vector currents. However, as discussed
in~\cite{FKS,OZI}, they cannot be expressed as $U(\theta){\rm diag}[f_8, f_1]$ due to the $U(1)_A$ anomaly.
Thus, the following two mixing angle parametrization is adopted~\cite{FKS,OZI}
\begin{eqnarray}\label{os_ang1}
&& f^8_\eta=f_8 \cos\theta_8, \;  f^1_\eta= -f_1 \sin\theta_1, \nonumber\\
&& f^8_{\eta'}=f_8 \sin\theta_8, \;  f^1_{\eta'}= f_1 \cos\theta_1.
\end{eqnarray}
The parameters appearing in Eq.~(\ref{os_ang1}) are related to the basis parameters
$\alpha, f_q\equiv f_{n\bar{n}}$ and $f_s\equiv f_{s\bar{s}}$, characterizing
the quark-flavor mixing scheme as follows~\cite{FKS}:
\begin{eqnarray}\label{os_decay}
&& f^2_8=\frac{f^2_q + 2 f^2_s}{3},\;\; \theta_8=\alpha -\arctan\biggl(\frac{\sqrt{2}f_s}{f_q}\biggr),\nonumber\\
&& f^2_1= \frac{2f^2_q + f^2_s}{3},\;\; \theta_1=\alpha -\arctan\biggl(\frac{\sqrt{2}f_q}{f_s}\biggr).
\end{eqnarray}

Using the plus component $(\mu=+)$ of the currents, one can
calculate the decay constants.
The explicit formulae of pseudoscalar and vector meson decay constants
in quark-flavor basis are given by~\cite{CJ_99,Jaus91}
\begin{align}
  \begin{split} \label{eqn:Decayconstant}
       f_{P}&=\sqrt{6}\int^1_0 dx \int \frac{
	d^{2}\mathbf{k}_{\bot}}{8\pi^{3}}
             \frac{\Phi(x,\mathbf{k}_{\bot})}{\sqrt{\mathcal{A}^{2}+\mathbf{k}_{\bot}^{2}}}
              \mathcal{A},\\
       f_{V}&=\sqrt{6}\int^1_0 dx \int \frac{d^{2}\mathbf{k}_{\bot}}{8\pi^{3}}
         \frac{\Phi(x,\mathbf{k}_{\bot})}{\sqrt{\mathcal{A}^{2}+\mathbf{k}_{\bot}^{2}}}
	 \left[\mathcal{A}+\frac{2\mathbf{k}_{\bot}^{2}}{D_{\rm LF}}\right],
     \end{split}
\end{align}
where $\mathcal{A}=(1-x) m_{q} + x m_{\bar{q}}$ and $D_{\rm LF}=M_{0}+m_{q}+m_{\bar{q}}$.


\section{Results and Discussion} 
\label{sec:ResultsandDiscussion}
\begin{figure*}
  \subfigure[]{\includegraphics[width=8.6cm]{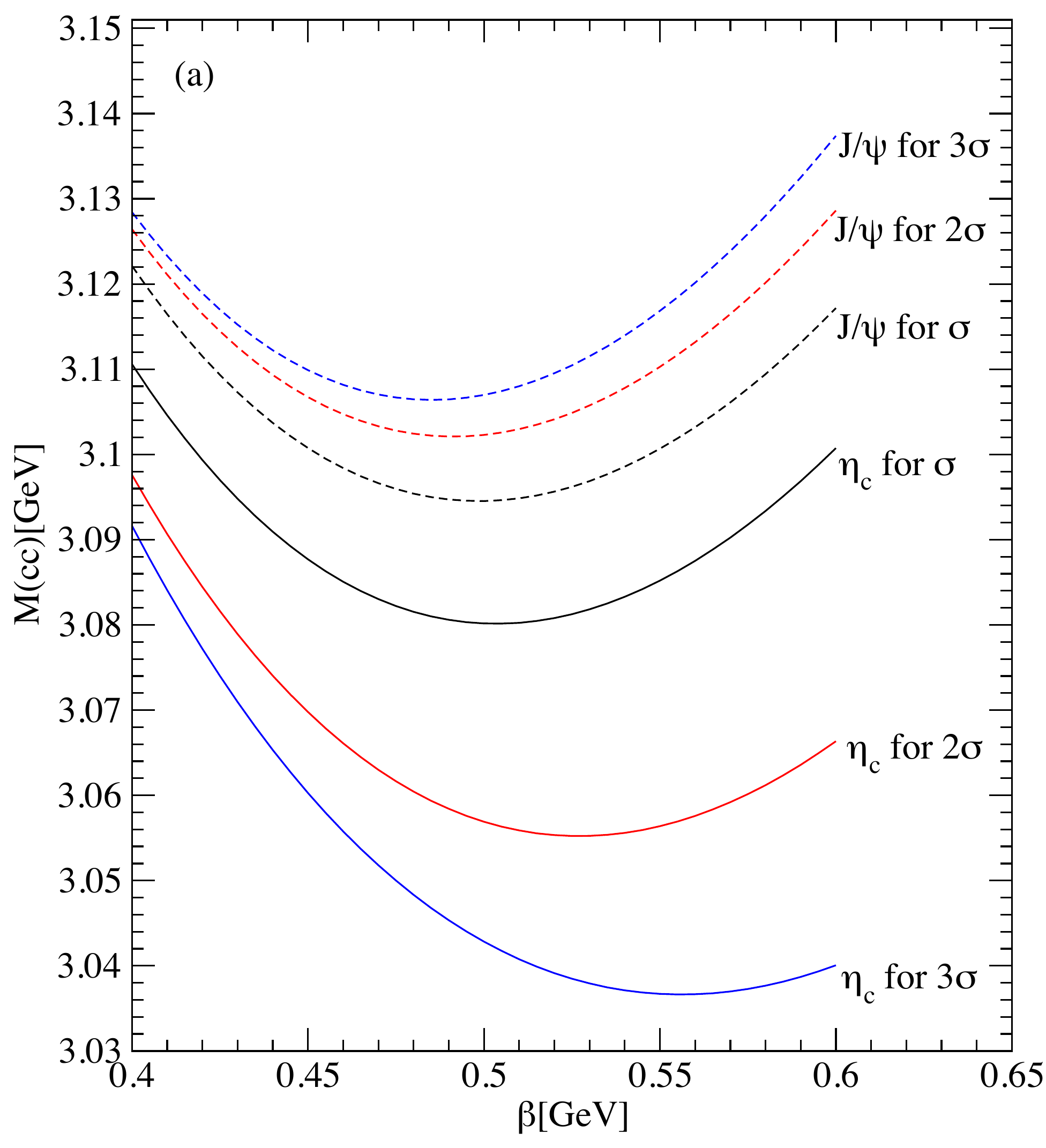}}
  \hspace{0.3cm}
  \subfigure[]{\includegraphics[width=8.6cm]{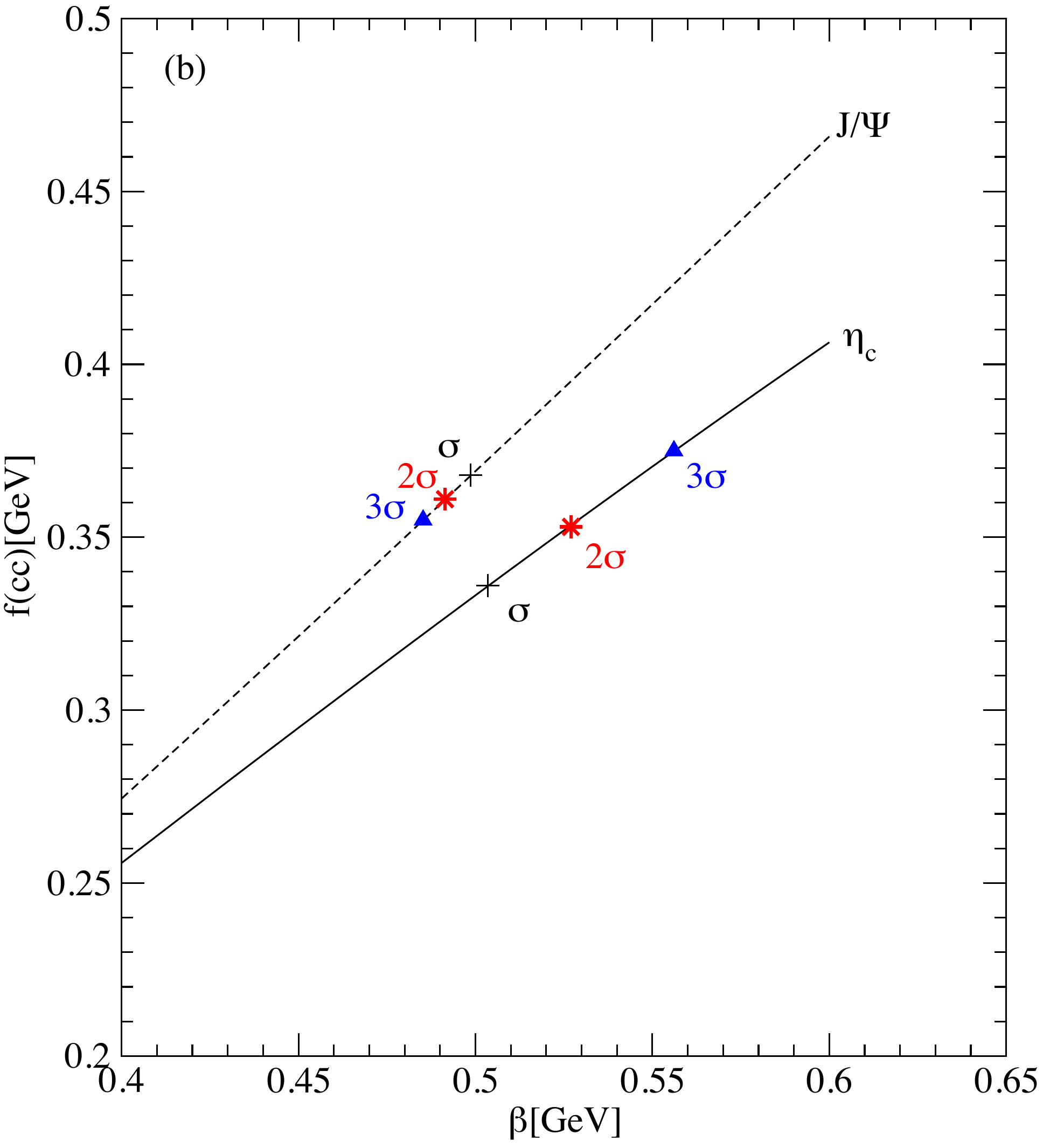}}
   \\
  \subfigure[]{\includegraphics[width=8.35cm]{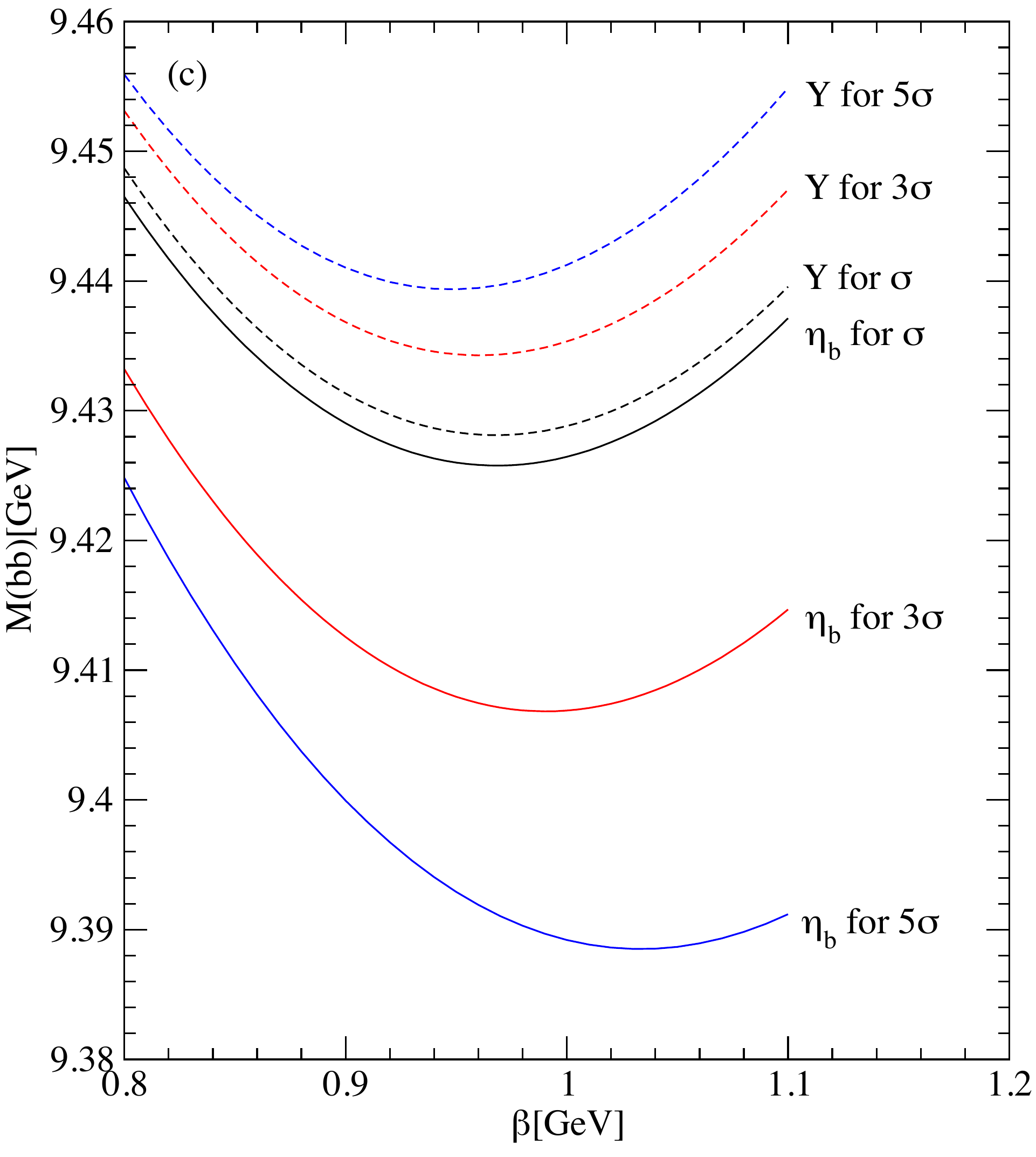}}
  \hspace{0.9cm}
  \subfigure[]{\includegraphics[width=8.35cm]{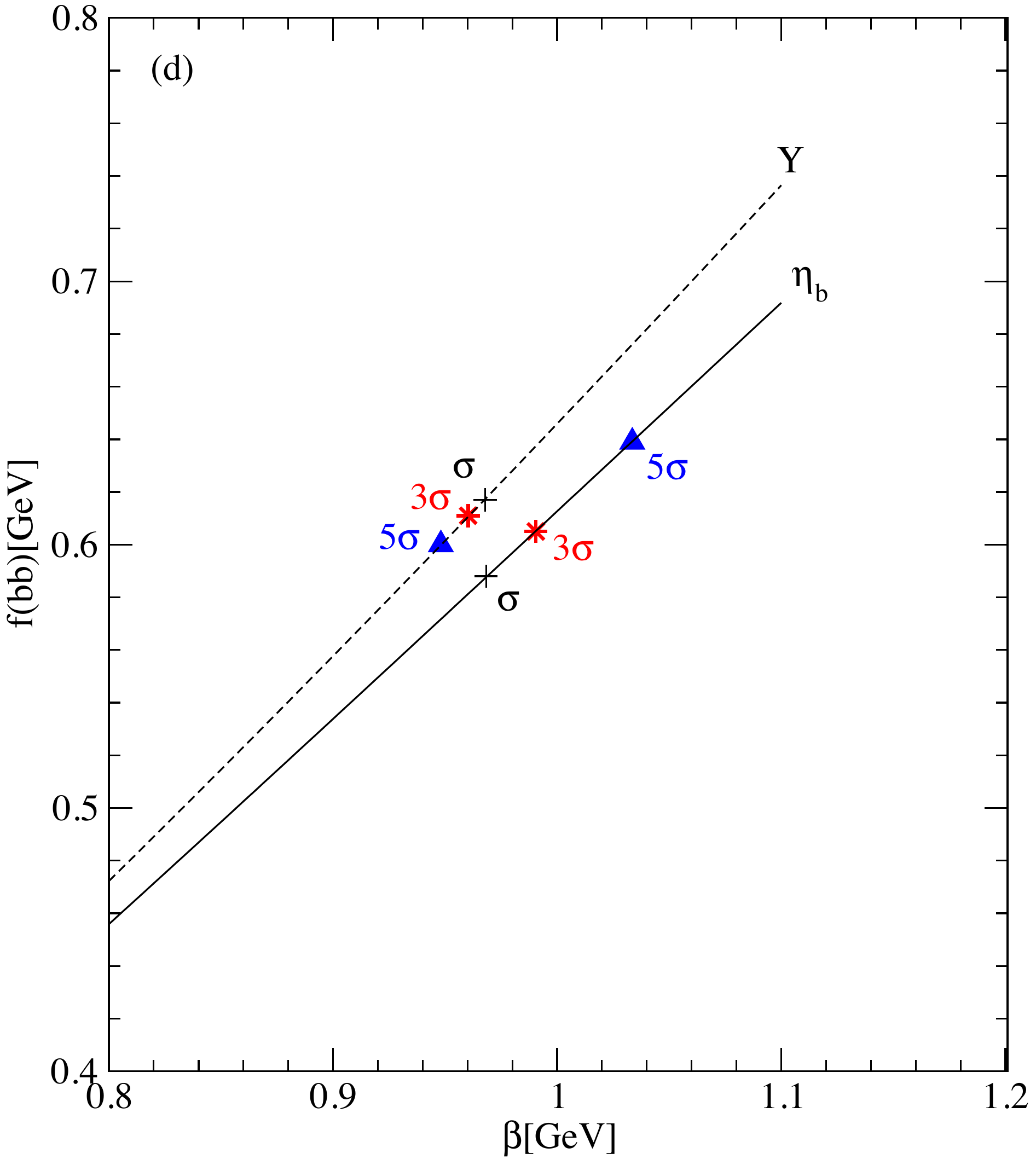}}
   \caption{\label{fig:Masshyper}(color online).
The masses (and hyperfine splittings) and the corresponding decay constants
of heavy quarkonia depending on the variation of the multiplicative factor $\lambda$,
i.e. $(c\bar{c})$ (Fig.~\ref{fig:Masshyper}(a) and~\ref{fig:Masshyper}(b))
with two different $\lambda\sigma=(1,2)\sigma$
and $(b\bar{b})$ (Fig.~\ref{fig:Masshyper}(c) and~\ref{fig:Masshyper}(d)) with two different $\lambda\sigma=(1,3)\sigma$
values, respectively.}
\end{figure*}


In Fig.~\ref{fig:Masshyper}, we show
the masses (and hyperfine splittings) and the corresponding decay constants
of heavy quarkonia depending on the variation of the multiplicative factor $\lambda$.
For $(c\bar{c})$ (Fig.~\ref{fig:Masshyper}(a) and~\ref{fig:Masshyper}(b))
and $(b\bar{b})$ (Fig.~\ref{fig:Masshyper}(c) and~\ref{fig:Masshyper}(d)) ,
we plot the curves corresponding
three different $\lambda$ values, i.e. $\lambda\sigma=(1,2,3)\sigma$
and $\lambda\sigma=(1,3,5)\sigma$, respectively.
The results indicate that the interaction between heavier quarks gets more point-like as
the larger $\lambda$ values are favored in comparison with data.
In general, as one can see from Fig.~\ref{fig:Masshyper},
the hyperfine splittings for both charmoninum (Fig.~\ref{fig:Masshyper}(a)) and
bottomonium (Fig.~\ref{fig:Masshyper}(c)) states increases as $\lambda$ increases
while other potential parameters remain the same.
On the other hand, the decay constants of vector mesons decrease while
the corresponding decay constants of pseudoscalar mesons increase as $\lambda$ increases
(see Fig.~\ref{fig:Masshyper}(b) and~\ref{fig:Masshyper}(d)).
One may increase $\lambda$ value even further to have better hyperfine splittings compared
to the data. However, one may not increase $\lambda$ value arbitrarily to accommodate
the empirical constraint $f_V\geq f_P$.
Similarly, we could improve the hyperfine splitting for $(b\bar{c}$) sector by using
$\lambda=2.3$.

\begin{figure}[!]
\begin{center}
  \includegraphics[width=8.6cm]{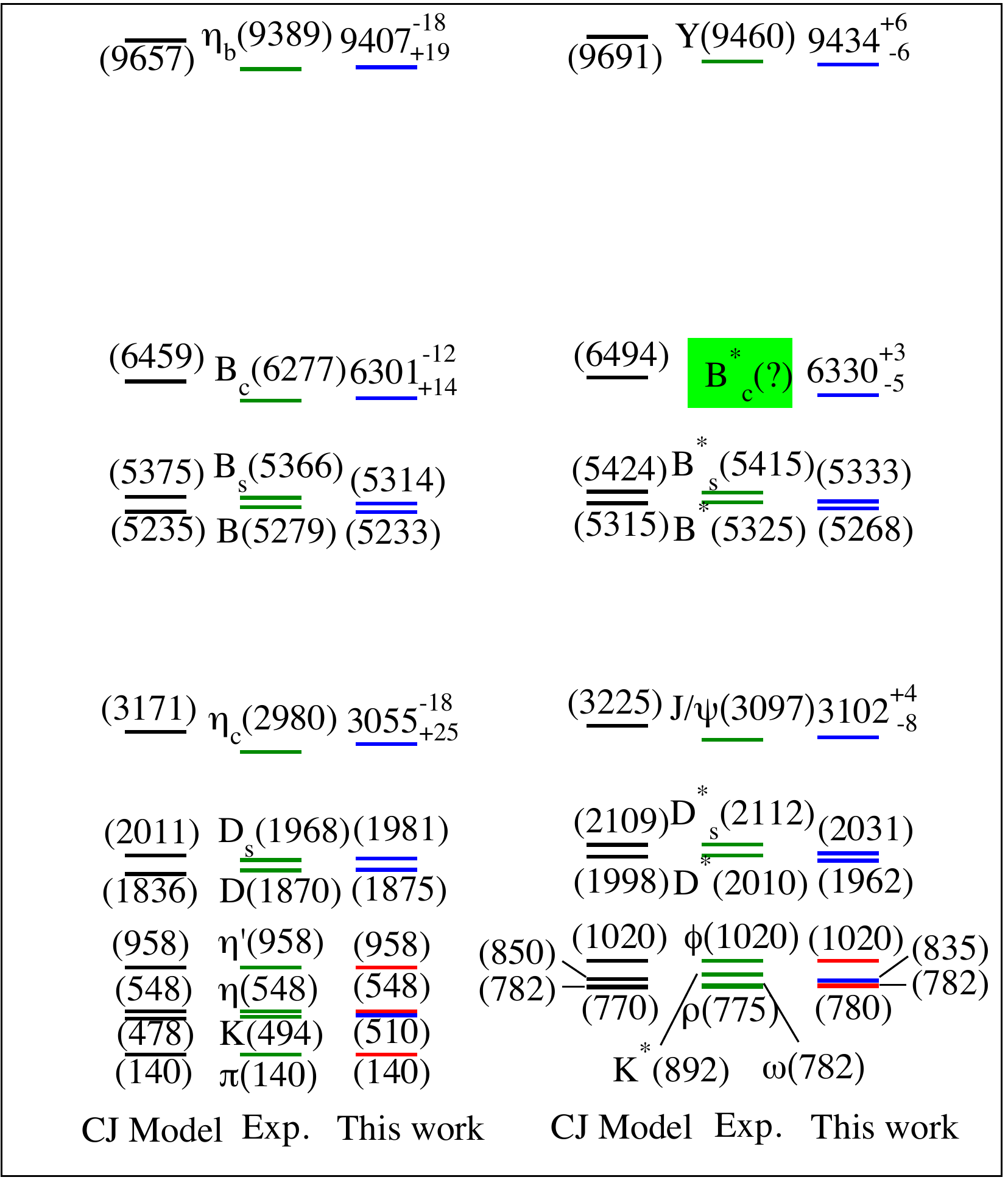}
  \caption{\label{fig:MassSpectrum}(color online). Fit of the ground state meson masses [MeV] with the parameters given
  in Table \ref{tab:betaP} and \ref{tab:quarkmass}  compared with the fit from our previous calculations using CJ model~\cite{CJ_Bc}
  as well as the experimental values.
  The $(\pi, \rho)$ masses are our input data. The $(\eta,\eta',\omega,\phi)$ masses are also used as input to find the $(\eta-\eta')$ and $(\omega-\phi)$
  mixing angles.  The theoretical error bars for $(c\bar{c}, b\bar{c}, b\bar{b})$ sectors are due to the usage
  of  $\lambda=(2^{+1}_{-1}, 2.3^{+1}_{-1}, 3^{+2}_{-2})$ values, respectively. }
\end{center}
\end{figure}

We show in Fig.~\ref{fig:MassSpectrum} our prediction of the meson mass spectra obtained from the variational principle
to the effective Hamiltonian with the smeared-out hyperfine interaction using the
trial function $\Phi=\sum_{i}^{2}c_i\phi_{iS}$  and compare them with the experimental data~\cite{PDG2014} as well as
the results obtained from the  CJ model with the linear confining potential~\cite{CJ_99}.
We should note that the $(\pi, \rho)$ masses are used as inputs.
The $(\eta,\eta',\omega,\phi)$ masses are also used as inputs to find the $(\eta-\eta')$ and $(\omega-\phi)$ mixing angles.
  The theoretical error bars for $(c\bar{c}, b\bar{c}, b\bar{b})$ sectors are due to the usage
  of  $\lambda=(2^{+1}_{-1}, 2.3^{+1}_{-1}, 3^{+2}_{-2})$ values, respectively.
As one can see, our trial wave function $\Phi$ including more HO basis
generates overall better results than our CJ model. This can be seen from our
$\chi^{2}=0.008$ compared to $\chi^{2}=0.012$
 obtained from the CJ model~\cite{CJ_Bc}.
Except the mass of $K$,
our predictions for the masses of $1S$-state pseudoscalar and vector mesons are within 4\% error.
Especially, our effective Hamiltonian
with the smeared hyperfine interaction using $\Phi$
clearly improves the predictions of heavy-light and heavy quarkonia systems
such as $(\eta_c, J/\psi, B_c, \eta_b, \Upsilon)$ compared to the CJ model adopting the contact hyperfine interaction.
Although the experimental data for $B^{*}_c$ is not yet available,
our predictions of $B^{*}_c$, i.e. $6330^{+3}_{-5}$ MeV,
are quite comparable with the lattice prediction $6331(9)$ MeV~\cite{HPQCD12} as well as other
quark model predictions such as  6340 MeV~\cite{GodfreyIsgurMeson} and 6345.8 MeV~\cite{FredPauli02}.

\begin{table}[!]
  \caption{\label{tab:lightMesonDecayConstants}Decay Constants for
light mesons (in unit of MeV) obtained from our updated LFQM.}
\renewcommand{\tabcolsep}{0.08pc}
    \begin{tabular}{lcccc}
				\hline
       Model & $f_{\pi }$ & $f_{\rho }$ & $f_K$ & $f_{K^*}$ \\[4pt]
    \hline
    This work & 130 &
      205 & 161 & 224 \\
      CJ~\cite{CJ_DA} & 130 & 246 & 161 & 256 \\
     \text{Exp}.~\cite{PDG2014}
     & \; $130.4(2)$ \;\;& \;$208^{(a)}, 216(5)^{(b)}$
     & \; $156.1(8)$\;\;&
     $217(7)$
     \\
		 \hline
    \end{tabular}
    $^{(a)}$ Exp. value for $\Gamma(\tau\to\rho\nu_\tau)$\\
    $^{(b)}$ Exp. value for $\rho^0\to e^+e^-$.
\end{table}

In Table~\ref{tab:lightMesonDecayConstants}, we list our predictions
for the decay constants of light mesons ($\pi,K,\rho,K^{*}$) obtained by using
the mixed wave function $\Phi$  of $1S$ and $2S$ HO states
and compare
them with the results from the CJ model~\cite{CJ_DA} and
the experimental data~\cite{PDG2014}.
As one can see, our updated model calculation including the hyperfine interaction in the variation procedure
clearly improves the results over the CJ model.

For the decay constant of $\phi$ meson, our prediction for the
ideal mixing angle~($\alpha^{\omega-\phi}_{\rm ideal}=90^\circ$) is given by
$f_\phi=f^{V}_{s\bar{s}}=245.1$ MeV. However, we obtain
$f_\phi=f^{V}_{s\bar{s}}=226$ MeV using our predicted mixing angle
$\alpha_{\omega-\phi}=84.8^\circ$.
Comparing to the experimental value $f^{\exp.}_\phi=233$ MeV~\cite{PDG2014}
(extracted from the partial width of $\phi\to e^+e^-$ decay), our prediction for
$f_\phi$ prefers a rather small $\omega-\phi$ mixing angle such as
$\alpha_{\omega-\phi}\simeq 87.5^\circ$
than the ideal mixing.

\begin{table}[!]
  \caption{\label{tab:mixingdecay}Decay constants in the singlet-octet basis and the mixing angle in the quark-flavor
  basis.}
\renewcommand{\tabcolsep}{0.5pc}
    \begin{tabular}{lccccc}
				\hline
      Reference & $f_8/f_{\pi }$ & $\theta_8$ & $f_1/f_{\pi}$ & $\theta_1$ & $\alpha$\\[4pt]
    \hline
    This work & 1.30 & $-27.3^\circ$ & 1.16 & $-8.6^\circ$ & $36.3^\circ$ \\
      ~\cite{FKS} & 1.26 & $-21.2^\circ$ & 1.17 &  $-9.2^\circ$ & $39.3^\circ$ \\
      ~\cite{Leut98} & 1.28 & $-20.5^\circ$ & 1.25 & $-4^\circ$ & $-$ \\
      ~\cite{EF} & 1.51 & $-23.8^\circ$ & 1.29 &  $-2.4^\circ$ & $40.7^\circ$ \\
      ~\cite{SSW} &1.27 & $-19.5^\circ$ & 1.17 &  $-5.5^\circ$ & $42.1^\circ$ \\
		 \hline
    \end{tabular}
\end{table}
For the decay constants of $\eta$ and $\eta'$, our predictions of the decay constants
$f_q$ and $f_s$
are given by $f_q=130$ MeV and $f_s=184.8$ MeV so that $f_q/f_\pi=1$ and $f_s/f_\pi=1.42$, where
the SU(3) breaking effect is manifest in the ratio $f_q/f_s\neq 1$.
Using Eq.~(\ref{os_decay}), we obtain
$f_8/f_\pi=1.30$ and $f_1/f_\pi=1.16$ with $\theta_8=-27.3^\circ$ and
$\theta_1=-8.6^\circ$, respectively.
In Table~\ref{tab:mixingdecay}, we compare our results for the decay constants in the singlet-octet basis
and the mixing angle in the quark-flavor basis with other theoretical
predictions~\cite{FKS,Leut98,EF,SSW}.
As one can see, our results
are consistent with other theoretical model results.
Since the experimental values are very well known for light mesons, this improvement is very encouraging.

In Table~\ref{tab:CharmMesonDecayConstants}, we list our predictions for the charmed meson decay constants $(f_{D},f_{D^{*}},f_{D_{s}},f_{D_{s}^{*}},f_{\eta_{c}},f_{J/\Psi})$ together with CJ model~\cite{Choi07}, lattice QCD~\cite{MesonDecayConstant24,DC25,BDKMS,BLS12}, QCD sum rules \cite{MesonDecayConstant26}, relativistic Bethe-Salpeter (BS) model \cite{MesonDecayConstant27}, relativized quark model \cite{MesonDecayConstant28}, and other relativistic quark model (RQM)~\cite{MesonDecayConstant29} predictions as well as the available experimental data~\cite{PDG2014,MesonDecayConstant32}.
We extract the experimental value $(f_{J/\Psi})_{\text{exp}}=(407 \pm 5)\;\text{MeV}$ from the data $\Gamma_{\text{exp}}(J/\Psi\rightarrow e^{+}e^{-})=(5.55\pm 0.14)\; \text{keV}$~\cite{PDG2014}
 and the formula
\begin{align}\label{eqn:DecayWidthtoDecayConstant}
  \Gamma(V \rightarrow
  e^{+}e^{-})=\frac{4\pi}{3}\alpha^{2}_{\rm QED} e^2_Q \frac{ f_{V}^{2} }{M_{V}},
\end{align}
where $e_Q$ is the electric charge of the heavy quark in units of $e$ (2/3 for $c$ and $-1/3$ for $b$).
We should note that our results of the ratios
$f_{D_{s}}/f_{D}=1.11$ and
$f_{\eta_{c}}/f_{J/\Psi}=0.98^{+0.08}_{-0.07}$
are quite comparable with the available experimental data,
$f_{D_{s}}/f_{D}=1.25\pm 0.06$ ~\cite{PDG2014} and
$f_{\eta_{c}}/f_{J/\Psi}=0.81\pm 0.19$~\cite{PDG2014,MesonDecayConstant32}, respectively.
Our result of the ratios
$f_{D_{s}^{*}}/f_{D^{*}}=1.13$
is also in good agreement with other theoretical model calculations such as $1.16 \pm 0.02\pm0.06$ from
the lattice QCD~\cite{BLS12} and
$1.10\pm 0.06$ from the BS model~\cite{MesonDecayConstant27}.

We list our results for the bottomed mesons $(f_{B},f_{B^{*}},f_{B_{s}},f_{B_{s}^{*}},f_{\eta_{b}},f_{\Upsilon})$ in Table \ref{tab:BottomMesonDecayConstants}, and compare with CJ model~\cite{Choi07}, lattice QCD~\cite{MesonDecayConstant24,DC25,CDDHL}, QCD sum rules \cite{MesonDecayConstant26}, BS model \cite{MesonDecayConstant27}, relativized quark model \cite{MesonDecayConstant28}, and RQM \cite{MesonDecayConstant29} predictions as well as the available experimental data \cite{PDG2014,MesonDecayConstant37}.
Note that we extract the experimental value $(f_{\Upsilon})_{\text{exp}}=(689\pm5)\; \text{MeV}$ from the data $\Gamma_{\text{exp}}(\Upsilon\rightarrow e^{+}e^{-})=1.340\pm 0.018\; \text{keV}$~\cite{PDG2014} and Eq.~(\ref{eqn:DecayWidthtoDecayConstant}) with $e^2_Q=1/9$ for $V=\Upsilon$.
Our results for the ratios $f_{B_{s}}/f_{B}=1.13$ and
$f_{B_{s}^{*}}/f_{B^{*}}=1.15$
are in good agreement with the  QCD sum rules~\cite{MesonDecayConstant26} predictions:
$f_{B_{s}}/f_{B}=1.17^{+0.04}_{-0.03}$,  and $f_{B_{s}^{*}}/f_{B^{*}}=1.20\pm 0.04$.
Ours are also in good agreement with the lattice results, $f_{B_{s}}/f_{B}=1.206(24)$~\cite{DC25}  and $f_{B_{s}^{*}}/f_{B^{*}}=1.17(4)^{+1}_{-3}$~\cite{MesonDecayConstant24}.
Our result for the ratio
$f_{\eta_{b}}/f_{\Upsilon}=0.99^{+0.07}_{-0.04}$
is consistent with the heavy quark symmetry $f_{\eta_{b}}/f_{\Upsilon} =1$ \cite{MesonDecayConstant38}.
One can also see that for heavy charmed and bottomed mesons, the trial wave function $\phi_{\rm B}$ produces better results when compared with the experimental data as well as the lattice results.

\begin{table*}[!]
  \caption{\label{tab:CharmMesonDecayConstants}Charmed meson decay
constants (in unit of MeV) obtained from our updated LFQM.
The theoretical error bars for $f_{\eta_c(J/\psi)}$ come from the variation of the smearing
parameters $\sigma$, i.e. $f_{\eta_c(J/\psi)}(2\sigma^{+\sigma}_{-\sigma})$.}

	\renewcommand{\tabcolsep}{0.2pc}
    \begin{tabular*}{\textwidth}{l @{\extracolsep{\fill}}cccccc}
				\hline
       Model & $f_D$ & $f_{D^*}$ & $f_{D_s}$ & $f_{D_s^*}$ & $f_{\eta_c}$ & $f_{J/\psi }$ \\[4pt]
    \hline
     This work & 208 & 230 & 231 & 260  & $353^{+22}_{-17}$ & $361^{-6}_{+7}$ \\
     CJ~\cite{Choi07} & 197 & 239 & 232 & 273 & 326 & 360 \\
     \text{Lattice} \cite{MesonDecayConstant24} & $211\pm 3\pm 17$ &
     $245\pm 20_{-2}^{+3}$ & $231\pm 12_{-1}^{+8}$ & $272\pm
     16_{-20}^{+3}$ & -- & -- \\
     \text{QCD}~\cite{DC25,BDKMS} & $208\pm 7$~\cite{DC25} & --
     & $250\pm 7$~\cite{DC25} & -- &  $387\pm 7$~\cite{BDKMS} & $418\pm 9$~\cite{BDKMS} \\
     \text{Sum}-\text{rules} \cite{MesonDecayConstant26}& $201^{+12}_{-23}$ & $242^{+20}_{-12}$
      & $238^{+13}_{-23}$ & $293^{+19}_{-14}$ & -- & -- \\
     \text{BS} \cite{MesonDecayConstant27} & $230\pm 25$ & $340\pm 23$
     & $248\pm 27$ & $375\pm 24$ & $292\pm 25$ & $459\pm 28$ \\
     \text{QM} \cite{MesonDecayConstant28} & $240\pm 20$ & -- &
     $290\pm 20$ & -- & -- & -- \\
     \text{RQM} \cite{MesonDecayConstant29} & 234 & 310 & 268 &
     315 & -- & -- \\
     \text{Exp}. & $206.7\pm 8.9$~\cite{PDG2014} & --
     & $257.5\pm 6.1$~\cite{PDG2014}
     & -- & $335\pm 75$~\cite{MesonDecayConstant32}
     & $407\pm 5$~\cite{PDG2014} \\[1pt]
		 \hline
    \end{tabular*}
\end{table*}

\begin{table*}[!]
  \caption{\label{tab:BottomMesonDecayConstants}Bottomed meson
decay constants (in unit of MeV) obtained from our updated
LFQM.
The theoretical error bars for $f_{\eta_b(\Upsilon)}$ come from the variation of the smearing
parameters $\sigma$, i.e. $f_{\eta_b(\Upsilon)}(3\sigma^{+2\sigma}_{-2\sigma})$.}
    \begin{tabular*}{\textwidth}{l @{\extracolsep{\fill}}cccccc}
				\hline
       Model & $f_B$ & $f_{B^*}$ & $f_{B_s}$ & $f_{B_s^*}$ & $f_{\eta
       _b}$ & $f_{\Upsilon }$ \\[4pt]
     \hline
     This work & 181 & 188 & 205 & 216
     & $605^{+32}_{-17}$ & $611^{-11}_{+6}$ \\
     CJ~\cite{Choi07} & 171 & 185 & 205 & 220 & 507 & 529 \\
     \text{Lattice} \cite{MesonDecayConstant24} & $179\pm
     18_{-9}^{+34}$ & $196\pm 24_{-2}^{+39}$ & $204\pm
     16_{-0}^{+41}$ & $229\pm 20_{-16}^{+41}$ & -- & -- \\
     \text{QCD}~ \cite{DC25,CDDHL} & $189\pm 8$~\cite{DC25} & -- &
     $228\pm 8$~\cite{DC25} & -- & -- & $649\pm 31$~\cite{CDDHL} \\
    \text{Sum}-\text{rules} \cite{MesonDecayConstant26} & $207^{+17}_{-9}$ & $210^{+10}_{-12}$
     & $242^{+17}_{-12}$ & $251^{+14}_{-16}$ & -- & -- \\
     \text{BS} \cite{MesonDecayConstant27} & $196\pm 29$ & $238\pm 18$
     & $216\pm 32$ & $272\pm 20$ & -- & $498\pm 20$ \\
     \text{QM} \cite{MesonDecayConstant28} & $155\pm 15$ & -- &
     $210\pm 20$ & -- & -- & -- \\
     \text{RQM} \cite{MesonDecayConstant29} & 189 & 219 & 218 &
     251 & -- & -- \\
     \text{Exp}. & $229_{-31-37}^{+36+34}$
     \cite{MesonDecayConstant37} & -- & -- & -- & -- &
     $689\pm 5$~\cite{PDG2014} \\[1pt]
		 \hline
    \end{tabular*}
\end{table*}

In Table \ref{tab:BcDecayConstant}, we present our model predictions for the decay constants of $f_{B_{c}}$ and $f_{B_{c}^{*}}$,
and compare them with other model calculations \cite{CJ_Bc,MesonBcDecayConstant5,MesonBcDecayConstant8,MesonBcDecayConstant42,MesonBcDecayConstant43,MesonBcDecayConstant44,
MesonBcDecayConstant45}. Our results are comparable with other model calculations.

\begin{table*}[!]
  \caption{\label{tab:BcDecayConstant} Bottom-charmed meson
decay constants(in unit of MeV) obtained from our updated LFQM.
The theoretical error bars for $f_{B_c{(B^*_c)}}$ come from the variation of the smearing
parameters $\sigma$, i.e. $f_{B_c{(B^*_c)}}(2.3\sigma^{+\sigma}_{-\sigma})$.}
    \begin{tabular*}{\textwidth}{l @{\extracolsep{\fill}}ccccccccc}
				\hline
      Model & This work & CJ~\cite{CJ_Bc}
      & \cite{MesonBcDecayConstant5}  &
      \cite{MesonBcDecayConstant8} & \cite{MesonBcDecayConstant42}
      & \cite{MesonBcDecayConstant43}  &
      \cite{MesonBcDecayConstant44}  &
      \cite{MesonBcDecayConstant45}  \\
    \hline
     $f_{B_c}$ & $389^{+16}_{-3}$ & 349 & 360 & 433 & 500 & $460 \pm\; 60$ & 517
     & $410 \pm\; 40$ \\
    \hline
     $f_{B^{*}_{c}}$ & $391^{-5}_{+4}$ & 369 & -- & 503 & 500 & $460 \pm\; 60$ & 517
     & -- \\
		 \hline
    \end{tabular*}
\end{table*}

\section{Summary and Conclusion} 
\label{sec:SummaryandConclusion}
In this work, we updated our LFQM by smearing out the Dirac delta function in the hyperfine interaction
to avoid the issue of negative infinity in applying the variational principle to the computation of meson mass spectra,
while our previous model(CJ model) used the perturbation method  to handle the delta function in the contact hyperfine interaction.
Using the mixed wave function
$\Phi$ of $1S$ and $2S$ HO states as the trial wave function,
we calculated both the mass spectra of the ground state pseudoscalar and
vector mesons and the decay constants of the corresponding mesons.
The flavor mixing effect has also been implemented for the meson systems of $(\omega,\phi)$ and $(\eta,\eta')$.

The variational analysis with $\Phi$ seems to improve the agreement with the data of
meson decay constants over the results of the CJ model. It also appears to provide the better agreement
with data in the heavy meson mass spectra.
Accommodating the empirical constraint $f_V\geq f_P$, we have shown that the mass spectra  and the hyperfine splittings
for heavy $(b,c)$ quark sector get improved by introducing the multiplicative factor $\lambda$ in front of the smearing parameter
$\sigma$, i.e. $\sigma \rightarrow \lambda\sigma$, in the smeared hyperfine interaction $(\sigma^3/\pi^{3/2})e^{-\sigma^{2}\mathbf{r}^{2}}$.
Our results indicate that the interaction between heavier quarks gets more point-like as
the larger $\lambda$ values are favored in comparison with data.
The distinction between the heavy meson sector and the light meson sector is rather
natural in our LFQM analyses.
To get more definite conclusion in this respect, further analysis
of other wave function related observables such as various meson elastic and transition form factors may be useful.
It may be also interesting to analyze radially excited meson states using the larger HO basis.

\begin{acknowledgements}
This work is supported by the US Department of Energy (No. DE-FG02-03ER41260).
The work of H.-M. Choi was supported in part by the National Research Foundation Grant funded by the Korean Government (NRF-2014R1A1A2057457) and that of Z. Li was supported in part by the JSA/Jefferson Lab fellowship.
\end{acknowledgements}

\appendix

\section{Fixation of the model parameters using variational
principle} 
\label{sec:AppendixFixingParameters}
In our model, we assumed SU(2) flavor symmetry and have the following parameters that need to be fixed: constituent quark masses $(m_{u(d)}, m_{s}, m_{c}, m_{b})$, potential parameters $(a, b, \alpha_{s})$, gaussian parameter $\beta$, and the smearing parameter $\sigma$.
For our
trial wave function $\Phi=\sum_{n=1}^2 c_n\phi_{nS}$, we also have the mixing factor $c_n (n=1,2)$ that we have to adjust.
Notice that the $\beta$ values here are not only different for different quark combinations, but also different for pseudoscalar and vector mesons of the same quark combination.
The reason for this is that the hyperfine interaction we included in our parameterization process gives different contributions to the masses of pseudoscalar and vector mesons and thus induces different parameterizations under variational principle.

We now illustrate our procedure for fixing these parameters.
The variational principle gives us one constraint:
\begin{align} \label{eqn:variationalprinciple}
  \pd{\langle \Phi|H|\Phi \rangle}{\beta}=\pd{M_{q
    \bar{q}}}{\beta}=0.
\end{align}
We can use this equation to rewrite the coupling constant $\alpha_{s}$ in terms of other parameters and plug it back into Eq.~(\ref{eqn:MassEquationMix}) and thus eliminate $\alpha_{s}$.
The string tension $b$ is fixed to be 0.18 GeV, a well known value from other quark model analysis \cite{GodfreyIsgurMeson,ScoraIsgur1995,IsgurScora1989}.
We will leave the quark masses and smearing parameter $\sigma$ and the mixing factor
$c_1$
as externally adjustable variables.
We picked a set of values for
$(m_{u(d)}, m_{s}, m_{c}, m_{b}, \sigma, c_1)$
and proceed with the following procedure to solve for the rest of parameters.

We are left with 3 more parameters $(a, \beta^{P}_{q\bar{q}} , \beta^{V}_{q\bar{q}})$ for mesons of a certain quark combination $(q\bar{q})$, where
$\beta^{P}_{q\bar{q}}$, $\beta^{V}_{q\bar{q}}$ are the gaussian parameters for pseudoscalar and
vector mesons, respectively.
Using the masses of $\pi$ and $\rho$ as our input values for $M_{q \bar{q}}$ in 
in Eq.~(\ref{eqn:MassEquationMix}), and the condition that our coupling constants $\alpha_{s}$ are the same for all these ground state pseudoscalar and vector mesons, we can fix the three model parameters $(a, \beta^{p}_{q\bar{q}} , \beta^{V}_{q\bar{q}})$
for $q=u$ or $d$ from the following three equations:
\begin{subequations}
  \begin{align}
    M_\pi(\beta^{P}_{q\bar{q}},a)&=0.140, \\
    M_\rho(\beta^{V}_{q\bar{q}},a)&=0.780, \\
    \alpha_{s}(\beta^{P}_{q\bar{q}},a)&=\alpha_{s}(\beta^{V}_{q\bar{q}},a).
  \end{align}
\end{subequations}
Solving these equations not only gives us the remaining parameters
$a, \beta^P_{q\bar{q}},\text{and} \beta^V_{q\bar{q}}$, but also the coupling constant $\alpha_{s}$ which we assumed to be the same for all the mesons we consider here.
We can then solve for the $\beta$ values of all the other mesons using the known $\alpha_{s}$ value, by equating the $\alpha_s$ expressions for different mesons that we got from Eq.~(\ref{eqn:variationalprinciple}).
We thus fixed all parameters for the ground state pseudoscalar and vector mesons we consider here.

We then assign a different set of values to the externally adjustable variables, i.e.
$(m_{u(d)}, m_{s}, m_{c}, m_{b}, \sigma, c_1)$,
and repeat the above procedure until we find a set of values that give best fit for the meson mass spectra.

Through our trial and error type of analysis, we found
$m_{q}=0.205\; \text{GeV}, m_{s}=0.38\; \text{GeV}, m_{c}=1.75\; \text{GeV}, m_{b}=5.15\; \text{GeV}, \sigma=0.423\; \text{GeV}, c_1=\sqrt{0.7}$
gives best fit.
We then determine the mixing angles from the mass spectra of $(\omega,\phi)$ and $(\eta,\eta')$
using Eqs.(\ref{Eq:mixang}) and (\ref{Eq:mixM}) as we have described in Sec.\ref{sec:ModelDescription}.
In addition, the multiplicative factor $\lambda$ in front of the smearing parameter $\sigma$ for the $(c\bar{c}, b\bar{c}, b\bar{b})$ systems were adjusted utilizing the hyperfine splittings of $(b,c)$ quark sectors as we have discussed in Sec.\ref{sec:ResultsandDiscussion}.
The updated $\beta$ values with this $\lambda$ adjustment are listed in Table \ref{tab:betaP}.


\begin{thebibliography}{54}


\bibitem{ConstituentQuarkfromQCD}
E.~Bagan \textit{et~al.\/},
\newblock Nuclear Physics B - Proceedings Supplements \textbf{54}, 208  (1997).


\bibitem{BoundStatesofQuarks1991}
W.~Lucha, F.~F. Sch{\"o}berl, and D.~Gromes,
\newblock Phys. Rept. \textbf{200}, 127  (1991).

\bibitem{OneGluonExchange}
A.~De~R\'ujula, H.~Georgi, and S.~L. Glashow,
\newblock Phys. Rev. D \textbf{12}, 147 (1975).


\bibitem{Kon} L. A. Kondratyuk and D. V. Tchekin, Physics of Atomic Nuclei {\bf 64}, 727 (2001).

\bibitem{Cheng97} H.-Y. Cheng, C.-Y. Cheung, and C.-W. Hwang,
\Journal{\PRD}{55}{1559}{1997}.

\bibitem{Hwang10} C.-W. Hwang, \Journal{\PRD}{81}{114024}{2010}.


\bibitem{CCP} P. L. Chung, F. Coester, and W. N. Polyzou,
\Journal{\PLB}{205}{545}{1988}.

\bibitem{Card95} F. Cardarelli, I.L. Grach, I.M. Narodetskii, G. Salme,
S. Simula, \Journal{\PLB}{349}{393}{1995}.




\bibitem{DiracForms}
P.~A.~M. Dirac,
\newblock Rev. Mod. Phys. \textbf{21}, 392 (1949).


\bibitem{BPP} S. J. Brodsky, H. -C. Pauli, and S. Pinsky, Phys. Rep. {\bf 301}, 299 (1998).


\bibitem{Zero1} S. J. Brodsky and D. S. Hwang, \Journal{\NPB}{543}{239}{1999}.

\bibitem{Zero2} J.P.B.C. de Melo, J.H.O. Sales, T. Frederico, and P.U. Sauer,
\Journal{\NPA}{631}{574c}{1998}.

\bibitem{Zero3} H.-M. Choi and C.-R. Ji, \Journal{\PRD}{58}{071901(R)}{1998}.

\bibitem{CJ_99} H.-M. Choi and C.-R. Ji,
\Journal{\PRD}{59}{074015}{1999}; \Journal{\PLB}{460}{461}{1999}.

\bibitem{CJ_Bc} H.-M. Choi and C.-R. Ji, \Journal{\PRD}{80}{054016}{2009}.

\bibitem{CJ_DA} H.-M. Choi and C.-R. Ji,
\Journal{\PRD}{75}{034019}{2007}.

\bibitem{Jaus99} W. Jaus, \Journal{\PRD}{60}{054026}{1999}.

\bibitem{Jaus03} W. Jaus, \Journal{\PRD}{67}{094010}{2003}.

\bibitem{Cheng04} H.-Y. Cheng, C.-K. Chua, and C.-W. Hwang,
\Journal{\PRD}{69}{074025}{2004}.

\bibitem{CJ_PV} H.-M. Choi and C.-R. Ji,
\Journal{\NPA}{856}{95}{2011}; \Journal{\PLB}{696}{518}{2011}.


\bibitem{MF12} J.P.B.C. de Melo and T. Frederico, \Journal{\PLB}{708}{87}{2012}.

\bibitem{Jaus90} W. Jaus, \Journal{\PRD}{41}{3394}{1990}.

\bibitem{Choi07} H.-M. Choi, \Journal{\PRD}{75}{073016}{2007}.

\bibitem{many-body-H}  S. Adler and A. Davis,
\Journal{\NPB}{244}{469}{1984};
A. Le Yaouanc, L. Oliver, S. Ono, O. Pene, and J. Raynal,
\Journal{\PRD}{31}{137}{1985};
F. Llanes-Estrada, S. Cotanch, A. Szczepaniak, and E. Swanson,
\Journal{\PRC}{70}{035202}{2004}.

\bibitem{DS} C. D. Roberts, and A. G. Williams,
\newblock Prog. Part. Nucl. Phys. \textbf{33}, 477 (1994);
C. J. Burden, Lu Qian, C. D. Roberts, P. C. Tandy, and M. J. Thomson,
\Journal{\PRC}{55}{2649}{1997}.

\bibitem{Chris-Pauli} H.C. Pauli and J. Merkel, \Journal{\PRD}{55}{2486}{1997}


\bibitem{CJ_V14} H.-M. Choi and C.-R. Ji, \Journal{\PRD}{89}{033011}{2014};
 \Journal{\PRD}{91}{014018}{2015}.


\bibitem{Brodsky2006}
S.~J.Brodsky and G.~F. De Teramond,
\newblock Phys. Rev. Lett. \textbf{96}, 201601 (2006);
G.~F. De Teramond and S.~J.Brodsky,
\newblock Phys. Rev. Lett. \textbf{102}, 081601 (2009);
G.~F. De Teramond, H.~G. Dosch and S.~J.Brodsky,
\newblock Phys. Rev. D \textbf{87}, 075005 (2013);
S.~J.Brodsky, G.~F. De Teramond, A. Deur and H.~G. Dosch,
\newblock Few-Body Syst. \textbf{56}, 621 (2015).

\bibitem{ChoiJi2008}
H.-M. Choi and C.-R. Ji,
\newblock Phys. Rev. D \textbf{77}, 113004 (2008).

\bibitem{Brodsky2008}
S.~J.Brodsky and G.~F. De Teramond,
\newblock Phys. Rev. D \textbf{77}, 056007 (2008).






\bibitem{IsgurCapstickBaryon}
S.~Capstick and N.~Isgur,
\newblock Phys. Rev. D \textbf{34}, 2809 (1986).

\bibitem{GodfreyIsgurMeson}
S.~Godfrey and N.~Isgur,
\newblock Phys. Rev. D \textbf{32}, 189 (1985).

\bibitem{Jaus91} W. Jaus, \Journal{\PRD}{44}{2851}{1991}.

\bibitem{Brodsky1977}
S.~J. Brodsky and G. P. Lepage,
\newblock SLAC-PUB-1966, Invited talk presented to the Fourth International Colloquium
on Advanced Computing Methods in Theoretical Physics, Saint Maximin, France, March 21-23, 1977.

\bibitem{Adkins1998}
G. S. Adkins and J. Sapirstein,
\newblock Phys. Rev. A \textbf{58}, 3552 (1998).

\bibitem{Barbieri1978}
R. Barbieri and E. Remiddi,
\newblock Nucl. Phys. B \textbf{141}, 413 (1978).

\bibitem{Bethe1947}
H.~A. Bethe,
\newblock Phys. Rev. \textbf{72}, 339 (1947).

\bibitem{Jones1997}
B. Jones, R. Perry, and S. G\l azek,
\newblock Phys. Rev. D \textbf{55}, 6561 (1997).


\bibitem{Lamm2014}
H. Lamm and R. F. Lebed,
\newblock J. Phys. G \textbf{41}, 125003 (2014).


\bibitem{ScoraIsgur1995}
D.~Scora and N.~Isgur,
\newblock Phys. Rev. D \textbf{52}, 2783 (1995).

\bibitem{BBD} A.M. Badalian, B.L.G. Bakker, and I.V. Danilkin,  Phys. At. Nucl. {\bf 74},  631 (2011).

\bibitem{FKS}  T. Feldmann, P. Kroll, and B. Stech,
\Journal{\PRD}{58}{114006}{1998}; \Journal{\PLB}{449}{339}{1999}.

\bibitem{OZI} T. Feldmann,  \Journal{\IJMPA}{15}{159}{2000}.

\bibitem{Leut98} H. Leutwyler,  Nucl. Phys. B (Proc. Suppl.) {\bf 64},  223 (1998).


\bibitem{PDG2014}
Particle Data Group, K.~A.~Olive \textit{et~al.\/},
\newblock Chin. Phys.  \textbf{C38}, 090001 (2014).

\bibitem{Scad} M. D. Scadron, \Journal{\PRD}{29}{2076}{1984}.

\bibitem{HPQCD12} R. Dowdall, C. Davies, T. Hammant, and R. Horgan (HPQCD collaboration),
\newblock  Phys. Rev. D \textbf{86}, 094510 (2012).

\bibitem{FredPauli02}
T.~Frederico, H.-C. Pauli, and S.-G. Zhou,
\newblock Phys. Rev. D \textbf{66}, 116011 (2002).



\bibitem{EF} R. Escribano and J.-M. Frere, J. High Energy Phys. JHEP06, 029 (2005) (arXiv:hep-ph/0501072).

\bibitem{SSW} J. Schechter, A. Subbaraman, H. Weigel, Phys. Rev. D \textbf{48}, 339 (1993).


\bibitem{MesonDecayConstant24}
D.~Becirevic \textit{et~al.\/},
\newblock Phys. Rev. D \textbf{60}, 074501 (1999).

\bibitem{DC25}
S.~Aoki \textit{et~al.}, FLAG Working Group,
\newblock  Eur. Phys. J. C \textbf{74}, 2890 (2014).
\%newblock Phys. Rev. Lett. \textbf{95}, 122002 (2005).

\bibitem{BDKMS}
D.~Be{\v c}irevi{\' c}, G. Duplan{\v c}i{\' c}, B. Klajn, B. Meli{\' c}, and F.  Sanfilippo,
\newblock Nucl. Phys. B \textbf{883}, 306 (2014).

\bibitem{BLS12} D. Becirevic, V. Lubicz, F. Sanfilippo, S. Simula, and C. Tarantino,
\newblock J. High Energy Phys. \textbf{02}, 042  (2012).

\bibitem{MesonDecayConstant26}
P. Gelhausen, A. Khodjamirian, A. A. Pivovarov, and D. Rosenthal,
\newblock Phys. Rev. D \textbf{88}, 014015 (2013); Phys. Rev. D \textbf{89}, 099901(E) (2014).

\bibitem{MesonDecayConstant27}
G.~Cveti{\v c}, C.~Kim, G.-L. Wang, and W.~Namgung,
\newblock Phys. Lett. B \textbf{596}, 84  (2004).

\bibitem{MesonDecayConstant28}
S.~Capstick and S.~Godfrey,
\newblock Phys. Rev. D \textbf{41}, 2856 (1990).

\bibitem{MesonDecayConstant29}
D.~Ebert, R.~Faustov, and V.~Galkin,
\newblock Phys. Lett. B \textbf{635}, 93  (2006).


\bibitem{MesonDecayConstant32}
(CLEO Collaboration), K.~W. Edwards \textit{et~al.\/},
\newblock Phys. Rev. Lett. \textbf{86}, 30 (2001).



\bibitem{CDDHL} B. Colquhoun, R. Dowdall, C. Davies, K. Hornbostel, and G. Lepage,
\newblock arXiv:1408.5768 [hep-lat].


\bibitem{MesonDecayConstant37}
Belle Collaboration, K.~Ikado \textit{et~al.\/},
\newblock Phys. Rev. Lett. \textbf{97}, 251802 (2006).

\bibitem{MesonDecayConstant38}
J.~P. Lansberg and T.~N. Pham,
\newblock Phys. Rev. D \textbf{75}, 017501 (2007).

\bibitem{MesonBcDecayConstant5}
M.~A. Ivanov, J.~G.~Korner, and P.~Santorelli,
\newblock Phys. Rev. D \textbf{63}, 074010 (2001), [hep-ph/0007169].

\bibitem{MesonBcDecayConstant8}
D.~Ebert, R.~N.~Faustov, and V.~O~Galkin,
\newblock Phys. Rev. D \textbf{67}, 014027 (2003), [hep-ph/0210381].

\bibitem{MesonBcDecayConstant42}
E.~J. Eichten and C.~Quigg,
\newblock Phys. Rev. D \textbf{49}, 5845 (1994), [hep-ph/9402210].

\bibitem{MesonBcDecayConstant43}
S.~S.~Gershtein,V.~Kiselev, A.~Likhoded, and A.~Tkabladze,
\newblock Phys. Rev. D \textbf{51}, 3613 (1995), [hep-ph/9406339].

\bibitem{MesonBcDecayConstant44}
L.~P. Fulcher,
\newblock Phys. Rev. D \textbf{60}, 074006 (1999), [hep-ph/9806444].

\bibitem{MesonBcDecayConstant45}
S.~Capstick and S.~Godfrey,
\newblock Phys. Rev. D \textbf{41}, 2856 (1990).

\bibitem{IsgurScora1989}
N.~Isgur, D.~Scora, B.~Grinstein, and M.~B. Wise,
\newblock Phys. Rev. D \textbf{39}, 799 (1989).


\end{thebibliography}

\end{document}